\documentclass[journal,comsoc,twocolumn]{IEEEtran}
\usepackage{amsthm,amsmath,amssymb}
\usepackage{mathrsfs}
\usepackage{booktabs}
\usepackage{siunitx}
\usepackage{amsfonts,amssymb}
\usepackage[T1]{fontenc}
\usepackage{hyperref}
\usepackage{makecell}
\usepackage{multirow}
\usepackage{diagbox}
\usepackage{cite,color}
\usepackage[pdftex]{graphicx}
\usepackage{graphicx, subfigure}
\usepackage{amsmath}
\IEEEoverridecommandlockouts
\usepackage[cmintegrals]{newtxmath}
\usepackage{threeparttable}

\usepackage[table,xcdraw]{xcolor}

\begin{document}

\title{AI-Driven Subcarrier-Level CQI Feedback}
\author{\IEEEauthorblockN{Chengyong~Jiang, \IEEEmembership{Graduate Student Member,~IEEE},
		Jiajia~Guo, \IEEEmembership{Member,~IEEE}, Yuqing~Hua,	\\ Chao-Kai~Wen, \IEEEmembership{Fellow,~IEEE},
		and Shi~Jin, \IEEEmembership{Fellow,~IEEE}
	}

    \thanks{C.~Jiang, J.~Guo, Y.~Hua, and S.~Jin are with the School of Information Science and Engineering, Southeast University, Nanjing, 210096, P. R. China. (e-mail: chengyongjiang@seu.edu.cn, jiajiaguo@seu.edu.cn, 220220803@seu.edu.cn, jinshi@seu.edu.cn). }
	\thanks{C.-K.~Wen is with the Institute of Communications Engineering, National Sun Yat-sen University, Kaohsiung 80424, Taiwan. (e-mail: chaokai.wen@mail.nsysu.edu.tw).}}

\maketitle
\begin{abstract}
The Channel Quality Indicator (CQI) is a fundamental component of channel state information (CSI) that enables adaptive modulation and coding by selecting the optimal modulation and coding scheme to meet a target block error rate. While AI‐enabled CSI feedback has achieved significant advances, especially in precoding matrix index feedback, AI‐based CQI feedback remains underexplored. Conventional subband‐based CQI approaches, due to coarse granularity, often fail to capture fine frequency‐selective variations and thus lead to suboptimal resource allocation.
In this paper, we propose an AI‐driven subcarrier‐level CQI feedback framework tailored for 6G and NextG systems. First, we introduce CQInet, an autoencoder‐based scheme that compresses per‐subcarrier CQI at the user equipment and reconstructs it at the base station, significantly reducing feedback overhead without compromising CQI accuracy. Simulation results show that CQInet increases the effective data rate by 7.6\% relative to traditional subband CQI under equivalent feedback overhead. Building on this, we develop SR‐CQInet, which leverages super‐resolution to infer fine‐grained subcarrier CQI from sparsely reported CSI reference signals (CSI‐RS). SR‐CQInet reduces CSI‐RS overhead to 3.5\% of CQInet’s requirements while maintaining comparable throughput.
These results demonstrate that AI‐driven subcarrier‐level CQI feedback can substantially enhance spectral efficiency and reliability in future wireless networks.

\end{abstract}


\IEEEpeerreviewmaketitle

\section{Introduction}
\label{introduction}

\IEEEPARstart{T}{he} evolution from fifth-generation (5G) to sixth-generation (6G) and Next-Generation (NextG) mobile networks is characterized by the deep integration of artificial intelligence (AI) into radio-access technologies \cite{union2022future,11098641,8786074}.  
Within this AI-enabled paradigm, {channel state information} (CSI) feedback is widely regarded as a key physical-layer application \cite{guo2022overview,8663966}.  
A rich body of work demonstrates that AI-based CSI-feedback schemes can compress and reconstruct CSI more accurately and with lower overhead than conventional techniques \cite{8322184,10746627,wang2021compressive,fjc}.  
The 3rd Generation Partnership Project (3GPP) has formally embraced this direction in the study item ``Study on Artificial Intelligence (AI)/Machine Learning (ML) for NR Air Interface'' \cite{213599,sum1146}, underscoring the strategic importance of AI-driven CSI feedback.

Most existing studies, however, concentrate on the {precoding matrix index} (PMI) \cite{guo2022overview}.  
By contrast, the {channel quality indicator} (CQI), which enumerates all available modulation-and-coding schemes (MCSs), remains comparatively underexplored.  
CQI is indispensable to adaptive modulation and coding (AMC) \cite{1638615,9024384,9148457}, whose goal is to select the highest-rate MCS that satisfies a target block-error-rate (BLER) constraint \cite{793310,9534770}.  
In 5G New Radio (NR), the 3GPP specification assigns a single CQI to each \emph{subband} of 24–96 subcarriers \cite{ts2020radio,8666153,10826731}.  
To obtain this subband CQI, the per-subcarrier signal-to-noise ratios (SNRs) are aggregated into an effective SNR (eSNR) via exponential or mutual-information mappings \cite{8684774,5699418,91247441}.  
However, because eSNR compresses heterogeneous frequency responses into a single scalar, it can misrepresent highly frequency-selective channels, thereby degrading downlink throughput.
 
To enhance communication performance, prior research has explored the integration of AI into the existing CQI selection
and feedback mechanism.  
For instance, a supervised learning method in \cite{8461864} maps subcarrier SNRs to BLER, \cite{8580924} combines an autoencoder with a CQI classifier, \cite{10024770} predicts outdated CQI, and \cite{8417549} leverages spatial correlation to curb feedback.  
Nevertheless, these solutions still operate at the subband level. This restriction poses a significant drawback as it hinders the accurate capture of subcarrier-level channel variations. When channel gain fluctuates sharply across subcarrier, the coarse granularity of subband CQI proves inadequate in precisely reflecting the communication quality across all subcarriers within a subband, which may overestimate weak tones, inflate BLER, and waste spectral resources \cite{8666153}.

A natural remedy is \emph{subcarrier-level} CQI feedback, which assigns an individual CQI to every subcarrier.  
This fine-grained approach can offer a more detailed and accurate assessment of channel conditions compared to the subband-level CQI feedback, allowing the communication system to make more informed decisions regarding modulation, coding, and resource allocation \cite{sc}.
Subcarrier-level adaptation also aligns with several recent 3GPP RAN contributions. For example, LG Electronics (RP-250159 \cite{250159}) proposes dynamic carrier management based on anchor and capacity subcarriers; the RAN2 moderator summary (RP-250036 \cite{250036}) lists single-/multi-carrier and multi-band control-channel design as priority items; and Ericsson (RP-250084 \cite{250084}) tightens spectral-efficiency and reliability KPIs that mandate finer-grained link adaptation.  
Collectively, these inputs signal a consensus that future releases will require CQI reporting beyond today’s subband granularity.

Despite its merits, subcarrier-level CQI feedback has not been adopted in previous generations because transmitting a CQI for every subcarrier incurs prohibitive uplink overhead \cite{sc}.  
Moreover, dense CSI-reference-signal (CSI-RS) configurations exacerbate the problem.  
Although existing autoencoder-based CSI feedback schemes such as CsiNet compress CSI matrices effectively \cite{8322184}, the specific challenge of CQI-feedback overhead remains open.
To close this gap, we develop an AI-assisted subcarrier-level CQI feedback framework tailored to 6G/NextG systems.  
In contrast with the 5G subband approach, our method provides per-subcarrier CQI while containing overhead through learned compression.  
The main contributions are as follows:

\begin{itemize}
    \item \textbf{Fine-grained CQI formulation:} We introduce a subcarrier CQI approach that bypasses eSNR mapping and computes CQI directly for each subcarrier, enabling accurate tracking of frequency selectivity and delivering higher spectral efficiency than subband CQI. Simulation evaluations demonstrate its superiority over existing subband-level CQI approaches.
    \item \textbf{CQInet:} We propose CQInet, an autoencoder that compresses subcarrier CQI at the user equipment (UE) and reconstructs it at the base station (BS), reducing CQI feedback overhead without sacrificing accuracy. Simulations show a 7.6\% effective-rate gain over subband CQI under the same feedback budget.
    \item \textbf{SR-CQInet:} Based on CQInet and addressing the sparse CSI-RS configuration in current systems, we propose SR-CQInet, which leverages super-resolution (SR) technology to further reduce CSI-RS overhead. This design reduces CSI-RS overhead to 3.5\% of CQInet while maintaining comparable throughput.
\end{itemize}

The remainder of this paper is organized as follows. Section~\ref{s2} reviews the signal model and the 5G subband-CQI baseline.  
Section~\ref{s3} details the proposed subcarrier-level CQI schemes and the associated CQI feedback frameworks.  
Section~\ref{s4} presents simulation results, and Section~\ref{s5} concludes the paper.

\medskip

\textbf{Notations:} Matrices are denoted by uppercase boldface letters and vectors by lowercase boldface letters. $\lfloor \cdot \rfloor$ denotes the floor function, $\|\cdot\|_{2}$ the Frobenius norm, and $\mathbb{R}^{m\times n}$ and $\mathbb{C}^{m\times n}$ the real and complex vector spaces of dimension $m\times n$, respectively.

\section{System Model and Subband CQI Scheme}
\label{s2}

\subsection{Signal Model}

In alignment with current standards, we consider a single-user MIMO orthogonal frequency-division multiplexing (OFDM) downlink chain in a cellular network. The BS is equipped with $N_{\rm t}$ antennas, and the UE has $N_{\rm r}$ antennas. There are $N_{\rm c}$ subcarriers configured in the physical downlink shared channel (PDSCH), which are grouped into $J$ subbands. Accordingly, the subband index for subcarrier $n \in \{1, 2, \dots, N_{\rm c}\}$ is defined as $j(n) = \lfloor nJ / N_{\rm c} \rfloor \in \{1, 2, \dots, J\}$. For simplicity of notation, we often omit the argument $(n)$ from the subband index $j$ when their relationship does not need to be explicitly emphasized. A time slot for CQI feedback consists of 14 OFDM symbols.
 
On the UE side, the received signal for subcarrier $n$ is given by 
\begin{equation}
{\bf y}_n ={\bf H}_n{\bf w}_{n}{x}_n+{\bf z}_n, \label{eq1}
\end{equation}
where ${\bf y}_n \in \mathbb{C}^{N_{\rm r} \times 1}$ denotes the received signal, ${\bf H}_n \in \mathbb{C}^{N_{\rm r} \times N_{\rm t}}$ denotes the complex-valued channel coefficients in the frequency domain, ${\bf w}_n \in \mathbb{C}^{N_{\rm t} \times 1}$ denotes the subband-based precoding vector at the BS, $x_n \in \mathbb{C}$ represents the data symbol, and ${\bf z}_n \in \mathbb{C}^{N_{\rm r} \times 1}$ denotes the additive white Gaussian noise with variance $\sigma^2$. The channel is assumed to remain stable within each time slot.
Then, the SNR $\rho_n$ for the $n$-th subcarrier is computed at the UE as
\begin{equation}
\rho_n = \frac{\|{\bf H}_n {\bf w}_{n}\|_{2}^2}{\sigma ^2}.  \label{eq2}
\end{equation}

\subsection{Subband CQI Scheme} \label{offset}

For ease of exposition, we consider a single-stream procedure. This procedure can be extended to the multi-stream case by incorporating multiple precoding matrices per subcarrier.
The CQI selection process under the standard subband CQI scheme is illustrated as follows \cite{ts2020radio,10826731,11091459}: 

\begin{itemize}
    \item \emph{Channel estimation:} Initially, based on the CSI-RS transmitted by the BS, the UE obtains the subcarrier CSI $\{{\bf H}_n\}_{n=1}^{N_{\rm c}}$ through channel estimation.
    
    \item \emph{PMI selection:} The UE then selects the precoding matrices $\{{\bf w}_n\}_{n=1}^{N_{\rm c}}$ based on the estimated CSI using PMI selection codebooks \cite{10906315,ts2020radio} for each subband. In 5G NR, the UE and BS share a pre-configured codebook set that includes multiple candidate precoding matrices with associated binary index values (PMIs). After channel estimation, the UE selects the optimal precoding matrix from the codebook using criteria such as maximizing channel capacity or minimizing chordal distance \cite{1468321}, and feeds back the corresponding PMI via the uplink. Upon receiving the PMI, the BS retrieves the matching matrix and uses it for precoding in downlink transmission.
    
    \item \emph{SNR calculation and eSNR mapping:} After PMI selection, the subcarrier SNRs $\{\rho_n\}_{n=1}^{N_{\rm c}}$ are computed using \eqref{eq2}. These SNRs are then mapped to eSNRs $\{\rho^{\rm SB}_j\}_{j=1}^{J}$. The mapping involves passing the SNRs through an information metric function to evaluate information content per subcarrier. The average information content is then computed across each subband, and the inverse of the metric function is applied to derive the corresponding eSNR. In practice, a predefined SNR–eSNR lookup table is often used to simplify this process.
    
    \item \emph{BLER–eSNR–CQI table lookup:} Using the BLER–eSNR–CQI table, the BLERs $\epsilon^{\rm SB}_{j,k}$ for subband $j$ under each CQI level $k$ are determined based on the corresponding eSNRs. This table is generated from Monte Carlo simulations under AWGN conditions and lists the probability of unsuccessful transmission for different CQI levels and SNRs \cite{9352811}.

    \item \emph{CQI selection and feedback:} The subband CQI $k^{\rm SB}_j$ is selected as the largest $k$ such that the BLER requirement $\epsilon_{\rm th}$ is satisfied:
        \begin{equation} \label{sb1}
            k^{\rm SB}_j = \mathop{\arg\max}_{k} \left\{ k \left| ~\forall k \in {\cal K}: \epsilon^{\rm SB}_{j,k} \leq \epsilon_{\rm th} \right. \right\},
        \end{equation}  
    where ${\cal K}$ is the index range of CQI. The UE then sends the CQI set $\{k_1^{\rm SB}, \ldots, k_J^{\rm SB}\}$ to the BS. To reduce feedback overhead, CQI offset coding is adopted. Specifically, the feedback consists of a $C_1$-bit wideband CQI and $C_2$-bit subband CQI offsets. The wideband CQI represents the average across all subbands, while each offset indicates the difference from this average in the range ${[-2^{C_2-1}+1, \, 2^{C_2-1}]}$. Thus, the total feedback overhead is $C_1 + C_2J$, with typical values $C_1=4$ and $C_2=2$ as specified in the 3GPP standard \cite{ts2020radio}.
    
    \item \emph{MCS determination and downlink transmission:} Based on the CQI report, the BS determines the MCS, comprising the modulation order $M^{\rm SB}_j$ and coding rate $R^{\rm SB}_j$, and estimates the expected BLER $\epsilon^{\rm SB}_j$. To determine the MCS, a pre-defined $C_1$-bit CQI table lists candidate MCSs, where $C_1$ represents $2^{C_1}$ kinds of CQI. In this table, CQI ranging from ${\cal K} \in \{ 0,1,\dots,2^{C_1}-1 \}$ are each associated with a specific MCS, consisting of modulation schemes (e.g., quadrature phase shift keying (QPSK), 16 quadrature amplitude modulation (QAM), 64QAM) and low-density parity-check (LDPC) coding rates. Three 4-bit CQI tables are defined in the 3GPP specification \cite{ts2020radio} for various service scenarios and BLER thresholds. In this work, we use the first CQI table \cite[Table 5.2.2.1-2]{ts2020radio}, as detailed in Table~\ref{cqitable}, and adopt a BLER threshold of $\epsilon_{\rm th} = 0.1$ in \eqref{sb1} as the CQI selection criterion.

    
    The downlink data stream is then transmitted using the selected MCSs. The effective rate $R^{\rm SB}$, representing the expected number of successfully transmitted bits per time slot, is calculated as
        \begin{equation}
            R^{\rm SB} = \sum_{j=1}^{J} \left(\frac{N_{\rm c}}{J}\right) M^{\rm SB}_{j} R^{\rm SB}_{j} (1 - \epsilon^{\rm SB}_{j}). \label{sb2}
        \end{equation}

\end{itemize}

\section{AI for Subcarrier CQI Feedback}
\label{s3}
In this section, we first introduce the proposed subcarrier CQI scheme for 6G and NextG systems, designed to improve upon the subband CQI scheme in current 5G networks by enabling more granular CQI selection. Subsequently, to address the significant overhead inherent in this scheme, we present CQInet for reducing feedback overhead and further propose SR-CQInet to reduce reference signal overhead.

\begin{table}[t] 
\centering
\caption{CQI Table \cite[Table 5.2.2.1-2]{ts2020radio} \label{cqitable}}
\begin{tabular}{|c|c|c|}
\hline
\textbf{CQI Index} & \textbf{Modulation} & \textbf{LDPC Code Rate ($\times$ 1024)}\\
\hline
0 & \multicolumn{2}{c|}{\textbf{out of range}} \\
\hline
1 & QPSK & 78  \\
\hline
2 & QPSK & 120  \\
\hline
3 & QPSK & 193  \\
\hline
4 & QPSK & 308  \\
\hline
5 & QPSK & 449  \\
\hline
6 & QPSK & 602 \\
\hline
7 & 16QAM & 378  \\
\hline
8 & 16QAM & 490  \\
\hline
9 & 16QAM & 616  \\
\hline
10 & 64QAM & 466  \\
\hline
11 & 64QAM & 567  \\
\hline
12 & 64QAM & 666  \\
\hline
13 & 64QAM & 772  \\
\hline
14 & 64QAM & 873  \\
\hline
15 & 64QAM & 948  \\
\hline
\end{tabular}
\end{table}

\subsection{Subcarrier CQI Scheme}
\label{s2a} 

\begin{figure}[t]
  \centering
  \includegraphics[width=1\linewidth]{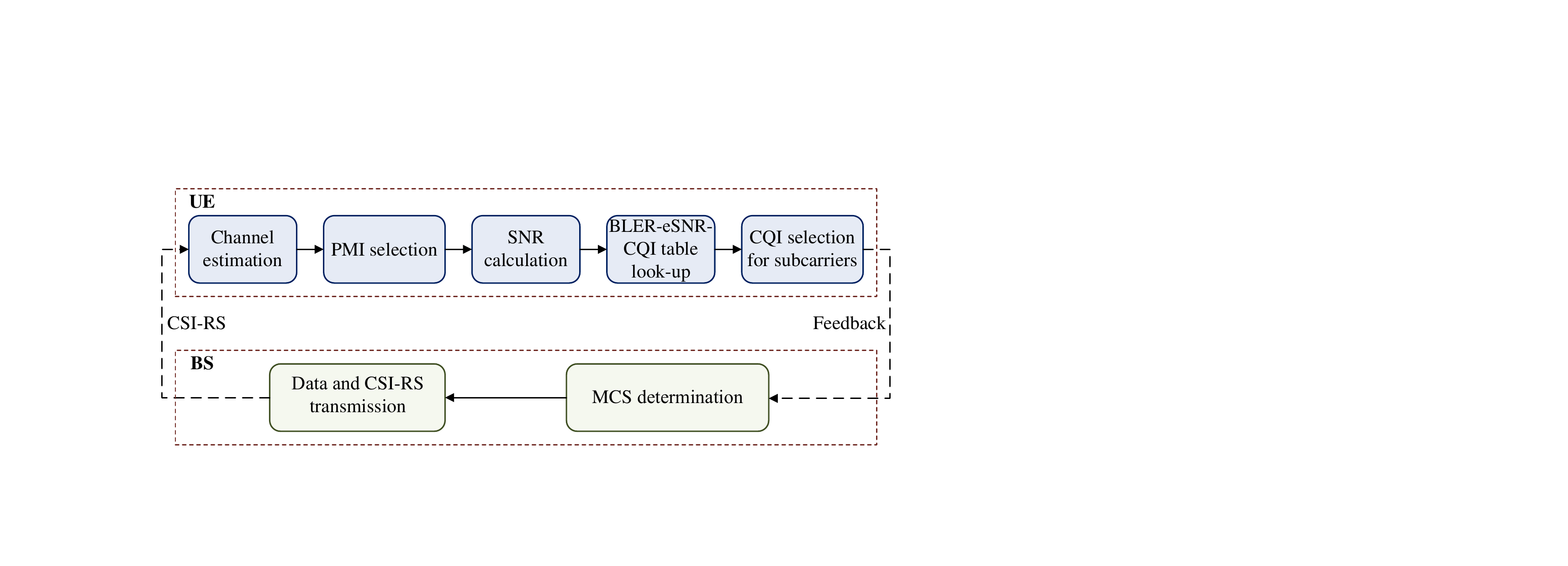}
  \caption{Schematic representation of the proposed subcarrier CQI scheme process.}
  \label{fig1}
\end{figure}

The procedure of the proposed subcarrier CQI scheme is illustrated in Fig.~\ref{fig1}. Both the $C_1$-bit CQI table and the BLER threshold $\epsilon_{\rm th}$ are predefined. As with the subband-based scheme, the UE performs channel estimation, subband-based PMI selection, and SNR calculation. Subsequently, the BLER $\epsilon_{n,k}$ for subcarrier $n$ and CQI level $k$ is derived directly from the BLER–eSNR–CQI table, without eSNR mapping. The selected subcarrier CQI $k_n$ for the $n$-th subcarrier is defined as 
\begin{equation} \label{eq3}
k_n = \mathop{\arg\max}_{k} \left\{ k \left| ~\forall k \in {\cal K}: \epsilon_{n,k} \leq \epsilon_{\rm th} \right. \right\}.
\end{equation}

Once the subcarrier CQI set $\{ k_1, \ldots, k_{N_{\rm c}} \}$ is reported to the BS, incurring a feedback overhead of $C_1 N_{\rm c}$, parameters such as the QAM order $M_n$, LDPC coding rate $R_n$, and expected BLER $\epsilon_n$ for each subcarrier are determined. The effective rate $R$ of the subcarrier CQI scheme is expressed as
\begin{equation}
R = \sum_{n=1}^{N_{\rm c}} M_n R_n (1 - \epsilon_n). \label{eq4}
\end{equation}

This subcarrier-centric approach introduces transformative capabilities for 6G and NextG systems. In contrast to the subband CQI scheme, which assigns a single CQI to a group of subcarriers, the proposed subcarrier CQI scheme individually selects the optimal CQI for each subcarrier, enabling the finest granularity in adaptation. It allows dynamic modulation and coding configuration based on the distinct channel quality of each subcarrier, rather than uniformly applying a single setting across a band.

This granularity is particularly critical in short-packet transmissions, where code blocks may span only a few dozen bits \cite{spt}. In such scenarios, traditional subband CQI’s coarse resolution may lead to mismatches, for instance, applying a high-rate MCS to a subcarrier with low SNR, resulting in degraded reliability. In contrast, subcarrier CQI ensures each subcarrier’s quality is precisely aligned with its payload, making it especially suitable for ultra-reliable low-latency communication (URLLC).

However, this increased flexibility incurs a cost: the feedback overhead scales linearly with the number of subcarriers $N_{\rm c}$. For comparison, the feedback overhead of the subband CQI scheme is $C_1 + C_2 J$, whereas for subcarrier CQI it is $C_1 N_{\rm c}$. This overhead difference becomes substantial in modern systems where $N_{\rm c}$ is large. For example, given $C_1 = 4$, $C_2 = 2$, $N_{\rm c} = 624$, and $J = 13$, the subcarrier CQI feedback overhead reaches $4 \times 624 = 2,496$ bits, while the subband CQI requires only $4 + 2 \times 13 = 30$ bits. Therefore, efficient CQI compression is indispensable for enabling practical deployment of the subcarrier CQI scheme.

\subsection{CQInet Framework}
In this subsection, we first introduce the motivation behind CQInet, followed by a detailed illustration of its architecture and training strategy. 

\subsubsection{Motivation}
An analysis of the subcarrier SNR sequence reveals that, despite fluctuations in subcarrier channel quality, the amplitude of these fluctuations remains small across several adjacent subcarriers. This phenomenon arises from the spatial correlation of wireless channels in OFDM systems, where adjacent subcarriers experience similar fading characteristics. Further observation of the subcarrier CQI sequence indicates that CQI values often remain constant across multiple adjacent subcarriers. These observations imply spatial redundancy in the CQI sequence, which can be exploited for compression.

On the other hand, due to the superior feature extraction capabilities of AI, neural networks (NNs), particularly autoencoder architectures, have demonstrated remarkable advantages in CSI feedback tasks \cite{guo2022overview}. In such tasks, the encoder extracts features from the channel matrix or precoding matrix and quantizes them into low-dimensional codewords, thereby reducing the CSI feedback overhead. The decoder then reconstructs the original matrix from the received codewords. This approach outperforms classical methods by adaptively learning intrinsic correlations rather than relying on predefined transforms.

Inspired by the successful application of autoencoders, we propose CQInet, a subcarrier-level CQI feedback algorithm leveraging an autoencoder-based architecture. CQInet compresses subcarrier CQI sequences by training NNs to learn intraband spatial correlations, thereby minimizing feedback overhead while reconstructing CQI values with high fidelity to preserve performance.

\subsubsection{CQInet Architecture}

\begin{figure}[t]
  \centering
  \includegraphics[width=1\linewidth]{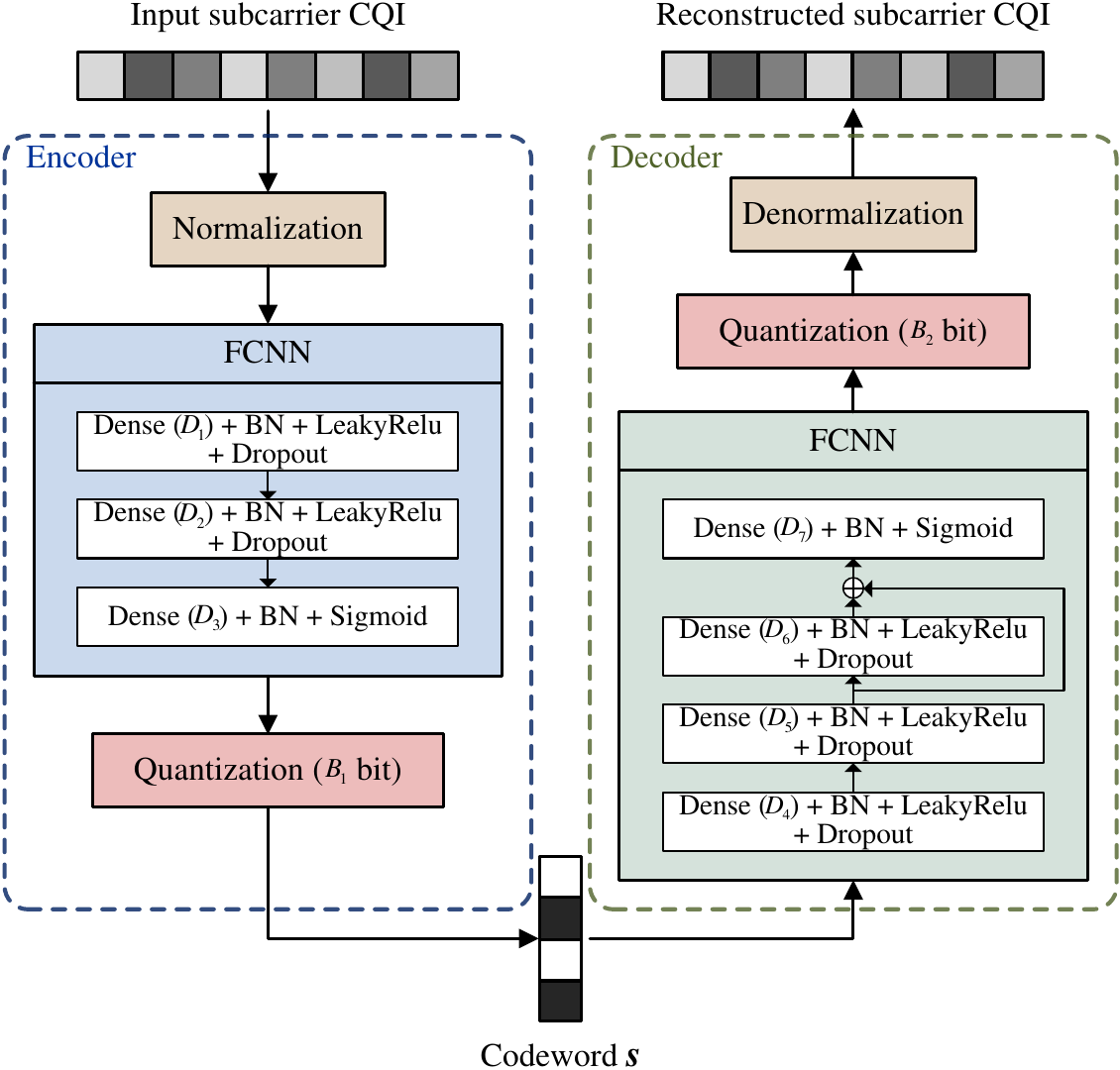}
  \caption{Schematic of the proposed CQInet architecture, utilizing an FCNN autoencoder for the compression and reconstruction of subcarrier CQI.}
  \label{fig2}
\end{figure}

The network architecture is depicted in Fig.~\ref{fig2}. We first consider a scenario where CSI-RSs are densely deployed across all subcarriers, enabling the UE to select the optimal subcarrier CQI ${\bf k} = \{ k_{n} \}^{N_{\rm c}}_{n=1}$ for all subcarriers based on the estimated CSI. The autoencoder in CQInet is then used to compress and reconstruct ${\bf k}$. The encoder resides at the UE and generates the codeword ${\bf s}$. Assuming error-free feedback, the decoder at the BS reconstructs the subcarrier CQI as $\hat{\bf k} = \{\hat{k}_{n}\}^{N_{\rm c}}_{n=1}$. A fully connected neural network (FCNN) is primarily adopted in this design. Accordingly, CQInet can be represented as 
\begin{align}
{\bf s} &= f_{\rm en}({\bf k};\, \boldsymbol{\Theta}),  \\
\hat{\bf k} &= f_{\rm de}({\bf s};\, \boldsymbol{\Phi}),
\end{align}
where ${\bf s} \in \mathbb{R}^{S \times 1}$ denotes the codeword of dimension $S = D_3 B_1$. Here, $D_3$ is the size of the last dense layer in the encoder, and $B_1$ is the quantization bit width. $f_{\rm en}(\cdot)$ and $f_{\rm de}(\cdot)$ denote the encoder and decoder functions, respectively, and $\boldsymbol{\Theta}$ and $\boldsymbol{\Phi}$ represent their parameters.

For the encoder, since each input CQI value $k_{n}$ is an integer within the CQI range ${\cal K}$, it is first normalized to facilitate training:
\begin{equation} \label{norm}
k_{n}^{\prime} = (k_{n}+0.5)/2^{C_1},
\end{equation}
ensuring the values lie within the range $[0, 1]$. The normalized value $k_{n}^{\prime}$ is then input to the FCNN. Each dense layer is followed by batch normalization (BN) to improve convergence. The first two dense layers use the LeakyReLU activation function to preserve gradient flow even for negative values, while the final layer uses a Sigmoid activation to bound the output within (0, 1), facilitating subsequent quantization. Dropout is applied after each dense layer except the last to mitigate overfitting.

The FCNN output is then passed through a quantization layer defined as 
\begin{equation} \label{eq5}
y = \left( \operatorname{round}(x \cdot 2^{B_1} - 0.5) + 0.5 \right) / 2^{B_1},
\end{equation}
where $x$ and $y$ are the input and output elements of the quantization layer. This uniformly quantizes $x$ into $2^{B_1}$ levels. The quantized output forms the codeword ${\bf s}$, with a resulting feedback overhead of $S = D_3 B_1$.

On the BS side, the decoder receives ${\bf s}$ as input to its FCNN. Similar to the encoder, BN and dropout are applied after each dense layer except the final one. The decoder is designed to be more complex than the encoder, leveraging the BS’s superior computational resources. To prevent gradient vanishing during training, a shortcut connection is added as suggested in \cite{he2016deep}. After decoding, the output is quantized using $B_2$ bits and denormalized via
\begin{equation} \label{de}
\hat{k}_{n} = \hat{k}_{n}^{\prime} \cdot 2^{C_1} - 0.5,
\end{equation}
which is the inverse of \eqref{norm}. The reconstructed output $\hat{k}_{n}$ thus lies within the valid CQI index range $\{0, 1, \dots, 2^{C_1}-1\}$.

\subsubsection{Training Setting}
\label{trainingsetting}
CQInet is trained offline using a simulated training dataset and then deployed online. During the training phase, the loss function of CQInet consists of two distinct components, each addressing a specific aspect of the model's performance.

First, the reconstructed CQI output by CQInet should closely approximate the ground-truth CQI input to minimize information loss during compression and reconstruction. Accordingly, the first component is the MSE between the target and reconstructed subcarrier CQI values, defined as 
\begin{equation}
{\cal L}_1 = \frac{1}{N_{\rm c}} \sum_{i=1}^{N_{\rm c}} \left( \hat{k}_{i} - k_{i} \right)^2.
\end{equation}
By minimizing ${\cal L}_1$, CQInet learns to preserve essential information in the CQI sequence throughout compression and reconstruction.

The second component addresses the asymmetry in the impact of CQI overestimation versus underestimation on system performance. Specifically, a one-level decrease in CQI only slightly reduces the effective data rate due to lower modulation and coding efficiency. In contrast, a one-level increase in CQI can significantly degrade performance, as the BLER may rise beyond 0.5, as reported in \cite{ts2020radio}. This asymmetry implies that CQI overestimation is substantially more harmful than underestimation.
To mitigate this issue, a penalty term is incorporated into the loss function to discourage CQI overestimation. This penalty term, denoted as ${\cal L}_2$, is defined as the average of the positive differences between the reconstructed and ground-truth CQI values:
\begin{equation}
{\cal L}_2 = \frac{1}{N_{\rm c}} \sum_{i=1}^{N_{\rm c}} \max \left( \hat{k}_{i} - k_{i}, 0 \right).
\end{equation}
Thus, when $\hat{k}_{i} > k_{i}$, a penalty is imposed. No penalty is applied if $\hat{k}_{i} \leq k_{i}$.

The overall loss function of CQInet is a weighted sum of the two components:
\begin{equation}
f_{\rm loss} = (1 - \alpha) \cdot {\cal L}_1 + \alpha \cdot {\cal L}_2,
\end{equation}
where $\alpha$ is a hyperparameter that controls the relative importance of the penalty term. By tuning $\alpha$, the training process can be adapted to prioritize either reconstruction accuracy or suppression of CQI overestimation, depending on the design requirements of the wireless communication system.
This combined loss function ensures that the model not only generates accurate reconstructions of the CQI sequence but also avoids overestimation, which is more detrimental to system reliability.


\subsection{SR-CQInet Framework}
In this subsection, we first introduce the motivation behind SR-CQInet and then describe its architecture. 

\subsubsection{Motivation}
The practical deployment of the CQInet algorithm faces challenges due to its reliance on densely deployed CSI-RS configurations to obtain accurate CQI for each subcarrier, which introduces significant overhead. In current systems, to reduce this overhead and improve resource utilization efficiency, CSI-RS is deployed sparsely, with signals transmitted only on a subset of subcarriers.
Given these operational constraints, dense CSI-RS configurations are unlikely to replace sparse configurations in the near future. Therefore, developing an algorithm capable of estimating fine-grained CQI from sparse channel measurements becomes essential for enabling subcarrier-level CQI feedback in future 6G and NextG systems.

SR techniques, originally developed in fields such as imaging and spectroscopy to enhance resolution beyond traditional limits \cite{10057379,94275031}, provide a natural framework for addressing this challenge. The core idea of SR is to recover detailed information from limited observations by exploiting underlying structures or data sparsity. Recently, deep learning-based SR methods have become mainstream, utilizing neural architectures to capture implicit priors and model non-linear mappings between low- and high-resolution domains.
In the context of wireless communications, several studies \cite{8370683,8400482,6817517,9427503} have employed SR techniques for channel estimation by leveraging spatial or frequency-domain correlations. For example, \cite{8370683} proposes an SR-based method for improving channel estimation accuracy in hybrid precoding systems with limited RF chains, leveraging the sparsity of mmWave channels. Similarly, \cite{8400482} exploits the sparse structure of massive MIMO channels to achieve SR-based channel estimation.

Motivated by these insights, we propose SR-CQInet, a subcarrier-level CQI feedback framework that integrates an autoencoder with an SR NN. To accommodate sparse CSI-RS deployment, the UE performs channel estimation and CQI calculation only on a limited subset of subcarriers, resulting in coarse-grained channel information. This coarse data is compressed into a low-dimensional codeword via an encoder. At the BS, a joint decoder and SR network reconstructs the full subcarrier-level CQI for MCS selection and downlink data transmission.
This architecture directly addresses the challenges of sparse CSI-RS deployment, enabling accurate CQI reconstruction from limited feedback. It reduces UE-side complexity and facilitates practical deployment, while maintaining the benefits of fine-grained, low-overhead CQI feedback.

\begin{figure}[t]
  \centering
  \includegraphics[width=1\linewidth]{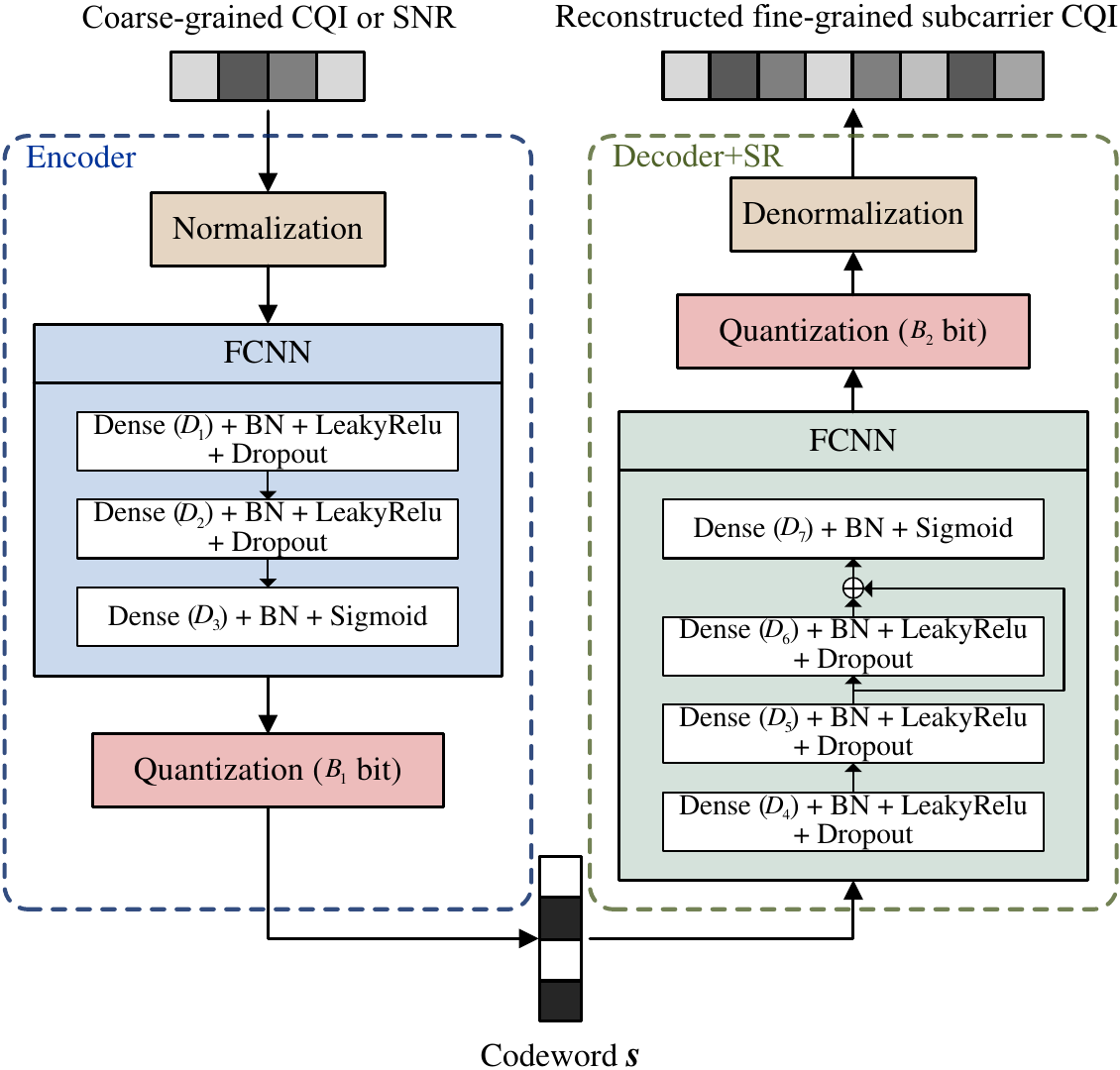}
  \caption{Schematic of the proposed SR-CQInet architecture, utilizing an FCNN autoencoder for the compression and reconstruction of subcarrier CQI.} 
  \label{srcqi}
\end{figure}

\subsubsection{SR-CQInet Architecture}
The network structure of SR-CQInet is illustrated in Fig.~\ref{srcqi}. Two types of input channel information are designed for SR-CQInet:

\begin{itemize}
    \item \textbf{Coarse-grained CQI:}  
    This type is derived from sparsely deployed CSI-RS and denoted as ${\bf k^{\rm CG}} = \{ k^{\rm CG}_{n} \}_{n=1}^{N_{\rm CG}}$, where $N_{\rm CG}$ is the number of subcarriers with CSI-RS across all $N_{\rm c}$ subcarriers. Since CQI is a discrete variable, reconstruction is relatively simpler, being limited to $2^{B_1}$ possible values. However, the limited information content may hinder SR performance due to insufficient representational richness.

    \item \textbf{Coarse-grained SNR:}  
    To mitigate the limitations of CQI discretization, we introduce a second input type, coarse-grained SNR values denoted as ${\boldsymbol{\rho}^{\rm CG}} = \{ \rho^{\rm CG}_{n} \}_{n=1}^{N_{\rm CG}}$, calculated from CSI-RS measurements. As continuous-valued inputs, SNR provides richer information, allowing the NN to better capture channel correlations for SR enhancement.
\end{itemize}

The operations of SR-CQInet are expressed as
\begin{align}
{\bf s}^{\rm sr} &= f_{\rm en}^{\rm sr}({\bf x}^{\rm CG};\, \boldsymbol{\Theta}^{\rm sr}), \\
\hat{\bf k}^{\rm sr} &= f_{\rm de\&sr}^{\rm sr}({\bf s}^{\rm sr};\, \boldsymbol{\Phi}^{\rm sr}),
\end{align}
where ${\bf x}^{\rm CG}$ denotes either ${\bf k^{\rm CG}}$ or ${\boldsymbol{\rho}^{\rm CG}}$.  
$f_{\rm en}^{\rm sr}(\cdot)$ and $f_{\rm de\&sr}^{\rm sr}(\cdot)$ represent the encoder and the joint decoder with SR functionality, respectively.  
$\boldsymbol{\Theta}^{\rm sr}$ and $\boldsymbol{\Phi}^{\rm sr}$ denote their corresponding network parameters.  
The resulting codeword ${\bf s}^{\rm sr} \in \mathbb{R}^{S \times 1}$ has dimension $S = D_3 B_1$, consistent with CQInet.  
The reconstructed fine-grained CQI is denoted as $\hat{\bf k}^{\rm sr} = \{ \hat{k}^{\rm sr}_n \}_{n=1}^{N_{\rm c}}$.
Notably, when using SNR as input, the network must learn the SNR-to-CQI mapping during training, making the task more challenging compared to using discrete CQI input.

Before inputting channel information to the encoder, normalization is performed to stabilize training. For CQI inputs, the normalization method from equation~\eqref{norm} is used. For SNR inputs, normalization is performed as
\begin{equation}
\rho_{n}^{\prime \mathrm{CG}} = \frac{\rho_{n}^{\mathrm{CG}}}{\max(\boldsymbol{\rho}^{\mathrm{CG}})}.
\end{equation}
The normalized coarse-grained input is processed by the FCNN encoder, where it is compressed into a low-dimensional feature vector through three dense layers. The output is then quantized using the same method as \eqref{eq5} to obtain the codeword ${\bf s}^{\rm sr}$, with quantization bit width $B_1$.

At the BS, the joint decoder and SR network processes ${\bf s}^{\rm sr}$ to reconstruct and upscale the output to size $N_{\rm c} \times 1$. This FCNN-based structure outputs the fine-grained subcarrier CQI $\hat{\bf k}^{\rm sr}$, which is then quantized with $B_2$ bits and denormalized using the same procedure as in CQInet.
Importantly, when using SNR as input, SR-CQInet can directly output subcarrier-level CQI, as the joint decoder and SR network implicitly learns the SNR-to-CQI mapping. This eliminates the need for additional computations at the BS.

To isolate the impact of the proposed SR-based design, core implementation components such as the quantization scheme and the base network structure at both the UE and BS follow the same conventions as CQInet. No additional explicit SR module is appended; instead, reconstruction and SR are jointly optimized. Training configurations remain consistent with those in Section~\ref{trainingsetting}, ensuring that performance differences are solely attributable to the super-resolution capability.

\section{Simulation Results} \label{s4}
In this section, we first describe the simulation settings for CQI generation and NN training. Then, we evaluate the proposed subcarrier CQI scheme, CQInet, and SR-CQInet, respectively. 

\begin{table}[t] \small
	\centering
	\caption{\label{tab1} Parameter setting.}
			\begin{tabular}{|c|c|}  \hline
			{\bf Parameter}  & {\bf Value}         \\
			\hline
			$ N_{\rm t} $  &  32     \\
			\hline
   $ N_{\rm r} $  &  4     \\
			\hline
Carrier frequency            & 2 GHz   \\
			\hline			
   Bandwidth            & 10 MHz   \\
			\hline
			{Subcarrier spacing}            & 15 KHz  \\
			\hline
			{Subband $J$}            & 13             \\
			\hline
			{Subcarrier $N_{\rm c}$}            & 624            \\
			\hline
			{Channel model}            & TDL-C  	\\
                \hline
                			{Delay spread}            & 300 ns  	\\
                \hline
			{Maximum Doppler shift}            & 50 Hz  	\\
                \hline

			{Average SNR}            & 5 dB
			\\ \hline
			{PMI codebook}            & Type I codebook \cite{ts2020radio}
			\\
			\hline
   
	\end{tabular}

\end{table}

\subsection{Simulation Settings}
The simulation parameters are summarized in Table~\ref{tab1}. Following the CQI selection procedure described in Section~\ref{s2}, we use the Type I codebook \cite{ts2020radio} to generate the precoding matrix ${\bf w}_{n}$ for subcarrier SNR computation.
The subcarrier CQI dataset is generated using the 5G NR downlink CSI reporting tool provided in MATLAB \cite{mat}. Our proposed methods are trained and evaluated using this dataset. The data is split into training, validation, and test sets in a 60\%-20\%-20\% ratio, with a total of 10,000 samples.
Notably, the SNR-to-eSNR mapping and the BLER–eSNR–CQI table look-up processes described in Section~\ref{s2} are implemented using the \texttt{hCQISelect} function from the 5G NR tool, which computes subband CQI directly from subcarrier SNR based on predefined CQI-versus-SNR lookup tables \cite{mat}.

To ensure a fair comparison between CQInet and SR-CQInet, we use consistent NN training configurations. The values of $C_1$ and $C_2$ are set to 4 and 2, respectively, in accordance with the 3GPP standard \cite{ts2020radio}. The quantization bit width $B_1$ is fixed at 2, while the width of the last dense layer $D_3$ is varied in the range $[5, 40]$ to generate different feedback overheads $S$ for evaluation.
The dimensions of the other dense layers are set as follows:
$D_1 = 300$, $D_2 = 100$, $D_4 = 100$, $D_5 = 300$, and $D_6 = D_7 = 624$.
A dropout rate of 0.03 is applied, and the penalty coefficient $\alpha$ in the loss function is set to 0.05. Training is conducted using the Adam optimizer for 1,000 epochs over 10,000 samples, with a batch size of 100. The learning rate is fixed at 0.001.

\subsection{Benchmark Performance for the Subcarrier CQI Scheme}
 
\begin{figure}[t]
  \centering
  \includegraphics[width=0.9\linewidth]{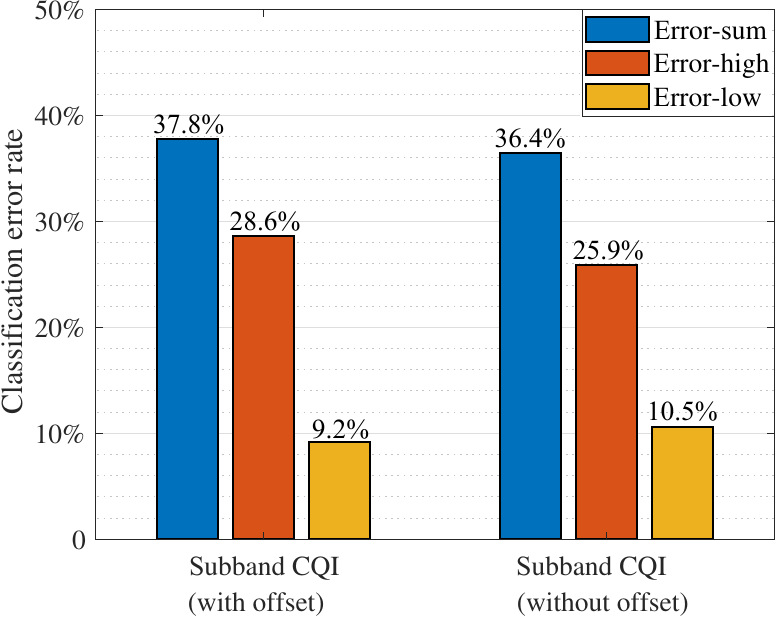}
  \caption{Comparison of classification error rates between subband CQI schemes with and without offset.}
  \label{fig4}
\end{figure}

The proposed subcarrier CQI scheme is evaluated against two benchmark methods:
\begin{itemize}
    \item \textbf{Subband CQI (with offset):}  
    This method corresponds to the standard CQI scheme described in Section~\ref{s2}, which computes subband-based CQI values and applies a 2-bit offset representation.  
    Although this approach significantly reduces feedback overhead compared to uncompressed subband CQI, the use of offsets limits the CQI range. As a result, the feedback may incur quantization errors, especially when the actual CQI deviations within a subband exceed the offset limits.  
    Under the simulation settings, the feedback overhead is fixed at $4 + 13 \times 2 = 30$ bits.

    \item \textbf{Subband CQI (without offset):}  
    This method eliminates the use of wideband CQI and offset encoding. Instead, each subband is directly represented using a full 4-bit CQI index, without any differential encoding.  
    This guarantees lossless CQI feedback at the cost of higher overhead. For 13 subbands, the total overhead becomes $13 \times 4 = 52$ bits.  
    This scheme serves as the upper performance bound for subband CQI feedback, offering complete fidelity in CQI reporting but without redundancy removal.
\end{itemize}

Fig.~\ref{fig4} shows the classification error rate of subband CQI schemes compared to subcarrier CQI. Since the CQI selection process can be interpreted as a classification task, the error types are categorized as ``Error-high'' (overestimation), ``Error-low'' (underestimation), and ``Error-sum'' (total misclassification rate).  
The cumulative error rate exceeds 36\%, with a large portion arising from CQI overestimation. High CQI overestimation leads to a significant increase in BLER, which severely degrades the effective throughput.  
Furthermore, the error rate of the subband CQI (with offset) is notably higher than that of the subband CQI (without offset) due to the constrained range of differential offsets. These constraints often fail to capture CQI deviations when the true offset is smaller than $-1$ or exceeds $2$.

\begin{figure}[t]
  \centering
  \includegraphics[width=0.9\linewidth]{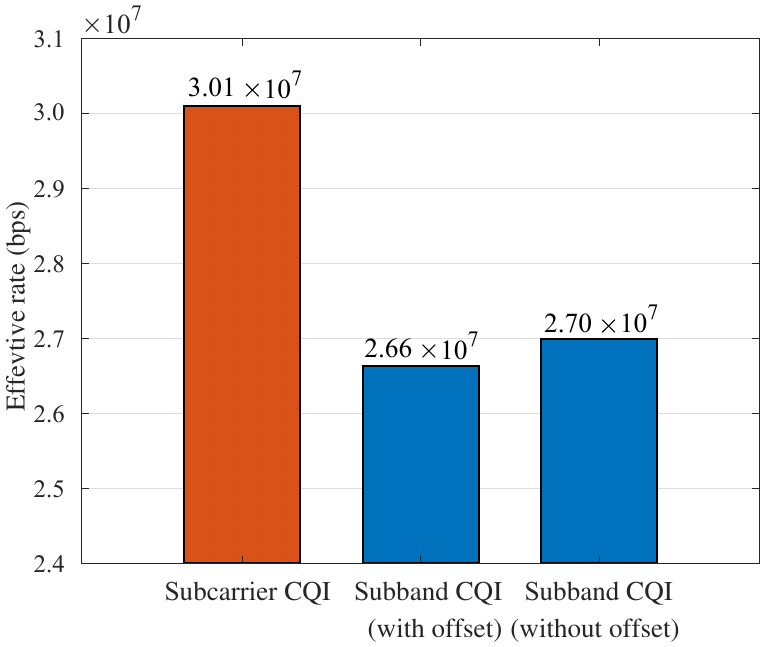}
  \caption{Comparison of effective rates between subcarrier and subband CQI schemes with and without offset.}
  \label{fig5}
\end{figure}

To evaluate the throughput performance of different schemes, Fig.~\ref{fig5} presents a comparison of their effective data rates. The proposed subcarrier CQI scheme achieves an effective rate of $3.01 \times 10^7$ bps, reflecting a $13.2\%$ gain over the subband CQI scheme with offset and an $11.5\%$ improvement over the scheme without offset.  
This performance advantage results from the subcarrier-level CQI feedback’s ability to capture fine-grained SNR variations, which facilitates more accurate MCS selection while mitigating BLER spikes commonly caused by CQI overestimation in coarse-grained schemes. These findings confirm the efficacy of subcarrier CQI feedback in 6G and NextG systems, where both high spectral efficiency and stringent BLER control are essential.
Moreover, Fig.~\ref{fig5} also confirms that the subband CQI scheme without offset performs better than its offset-based counterpart. This is attributed to the restrictive nature of offset encoding, which may force subband CQI values to fall outside the optimal range, resulting in either increased BLER or a more conservative selection of modulation and coding parameters.

\begin{figure}[t]
  \centering
  \includegraphics[width=0.9\linewidth]{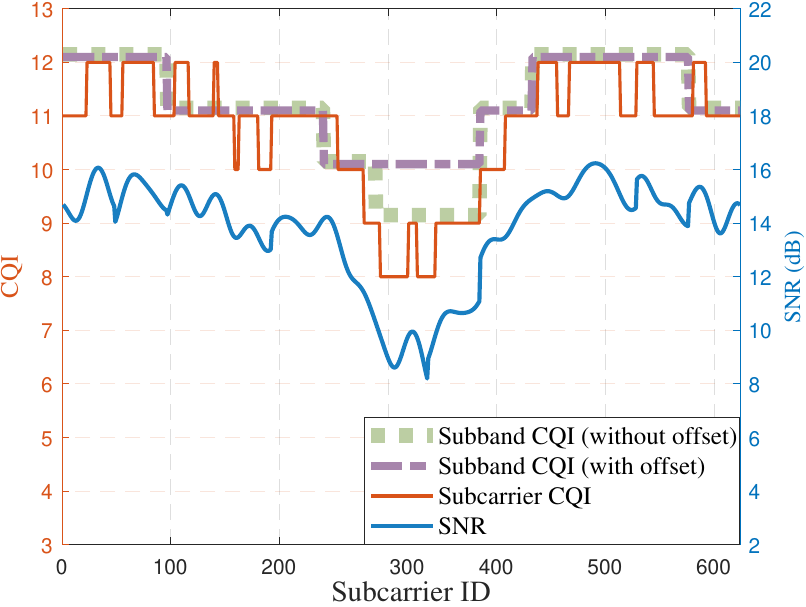}
  \caption{Sample comparison of ground-truth subcarrier CQI, proposed subcarrier CQI scheme, subband CQI schemes (with and without offset), and the corresponding subcarrier SNR.}
  \label{fig3}
\end{figure}

To visualize the CQI accuracy in practice, a subcarrier CQI sample is randomly selected, and its corresponding subcarrier SNR is plotted in Fig.~\ref{fig3}. Significant differences are observed between subcarrier and subband CQI schemes, particularly in regions with rapidly fluctuating SNR.  
For instance, between the 97th and 144th subcarriers, SNR varies sharply, resulting in subcarrier CQI values oscillating between 11 and 12. In contrast, the subband CQI remains fixed at 11, failing to reflect these variations.  
Additionally, from the 241st to 384th subcarriers, a valley-shaped SNR pattern emerges due to channel frequency selectivity. The offset-based subband CQI fails to accurately represent this region due to limited offset range, resulting in deviations of up to two levels between subband and subcarrier CQI.  
Even the unbiased subband CQI scheme, despite having full-range representation, suffers from averaging effects inherent to subband granularity, with deviations of one level in most cases.

These findings highlight the strength of the proposed subcarrier CQI scheme in capturing the frequency-selective nature of the channel. By providing CQI feedback at the subcarrier level, the scheme enables highly adaptive MCS selection that closely follows SNR trends, maximizing throughput in dynamic environments. In contrast, subband CQI schemes introduce quantization or averaging errors that degrade performance, especially in channels with fast frequency-domain variation.

\subsection{Performance of CQInet}

The proposed CQInet is evaluated against two benchmark methods. In addition to the previously introduced subband CQI scheme with fixed offset, we introduce the following enhanced scheme for comparison:
\begin{itemize}
    \item \textbf{Subband CQI (Variable Offset Size (VOS)):}  
    This enhanced scheme modifies the traditional subband CQI with fixed offset by varying the number of offset bits to better approximate the performance of subband CQI without offset, while reducing feedback overhead.  
    As suggested in \cite{b11}, although different subbands may require varying bit lengths based on the bit pattern, a uniform number of offset bits is used across all subbands in our evaluation to minimize overhead. Under the simulation settings, the average feedback overhead for this scheme is 34.8 bits.
\end{itemize}

\begin{figure}[t]
  \centering
  \includegraphics[width=0.9\linewidth]{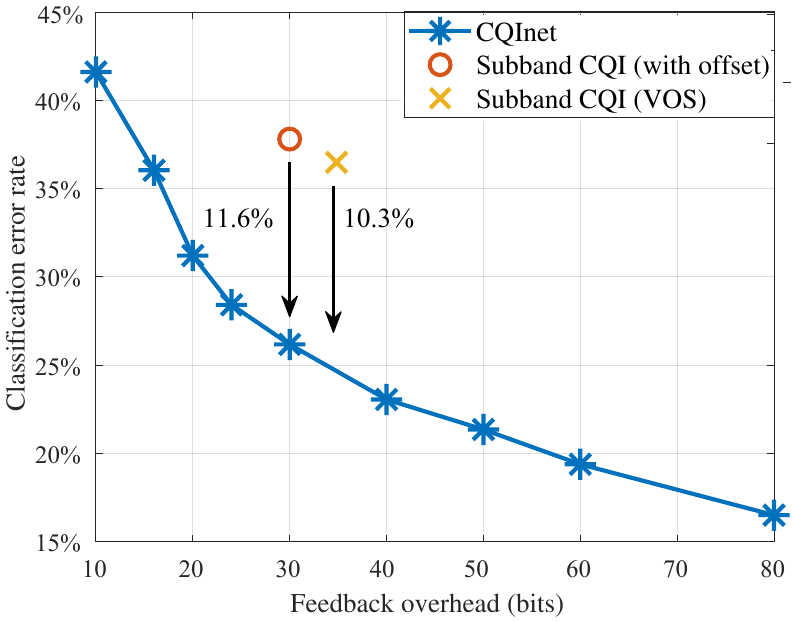}
  \caption{Comparison of classification error rates between CQInet and subband CQI schemes with fixed and variable offset.}
  \label{fig6}
\end{figure}

\begin{figure}[t]
  \centering
  \includegraphics[width=0.9\linewidth]{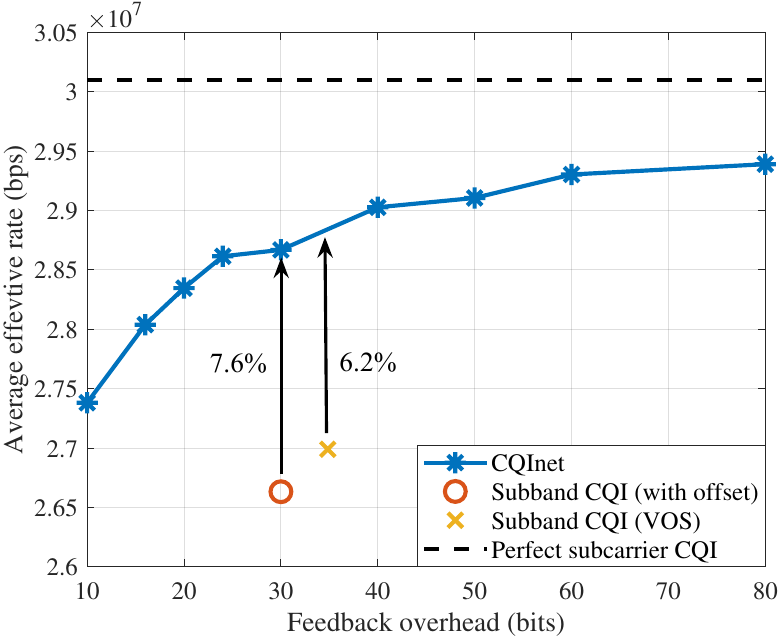}
  \caption{Comparison of effective rates between CQInet and subband CQI schemes with fixed and variable offset.}
  \label{fig7}
\end{figure}
 
Fig.~\ref{fig6} presents the classification error rates of CQInet and the benchmark schemes as a function of feedback overhead $S$. When compared with the subband CQI scheme using a fixed offset and under identical feedback overhead (30 bits), CQInet achieves an 11.6\% reduction in error rate. This demonstrates its superior capability in accurate CQI selection despite having the same overhead constraint.
Moreover, to match the accuracy level of the fixed-offset scheme, CQInet requires only about half the feedback overhead, showcasing its efficiency. Against the VOS-based scheme, CQInet delivers a 10.3\% lower error rate under the same overhead conditions. To achieve the same level of accuracy as VOS, CQInet reduces the overhead by approximately 60\%.
These results confirm that CQInet effectively exploits the correlation between adjacent subcarriers through its autoencoder-based architecture. As a result, it delivers high-accuracy, fine-grained CQI feedback with significantly reduced overhead compared to traditional subband-based methods.

Fig.~\ref{fig7} further compares the effective rates of these schemes under varying feedback overhead. When matched with the overhead levels of the fixed-offset and VOS schemes, CQInet achieves effective rate improvements of 7.6\% and 6.2\%, respectively.  
Notably, CQInet can attain these gains with only one-third of the original overhead, emphasizing its efficiency.
This superior performance arises from two key aspects. First, CQInet provides more accurate CQI feedback by leveraging learned spatial correlations. Second, it incorporates an overestimation penalty in its loss function, which discourages aggressive CQI prediction and thereby mitigates the risk of BLER escalation. This dual mechanism not only enhances link reliability but also improves the overall effective throughput.

Additionally, as feedback overhead increases, the effective rate of CQInet continues to improve, though with diminishing returns. Initially, the added overhead allows for finer CQI representation, resulting in better MCS selection. However, beyond a certain point, the marginal gains taper off as most relevant channel information is already captured, and further bits offer limited additional value.
In contrast to traditional subband CQI schemes, which operate under a fixed feedback configuration, CQInet demonstrates remarkable flexibility. It can dynamically adjust its feedback granularity and overhead to suit varying channel conditions and service requirements. This adaptability enhances system efficiency and enables intelligent, context-aware AMC, making CQInet a robust and scalable solution for future 6G and NextG networks.

\subsection{Performance of SR-CQInet}
\begin{figure}[t]
  \centering
  \includegraphics[width=0.9\linewidth]{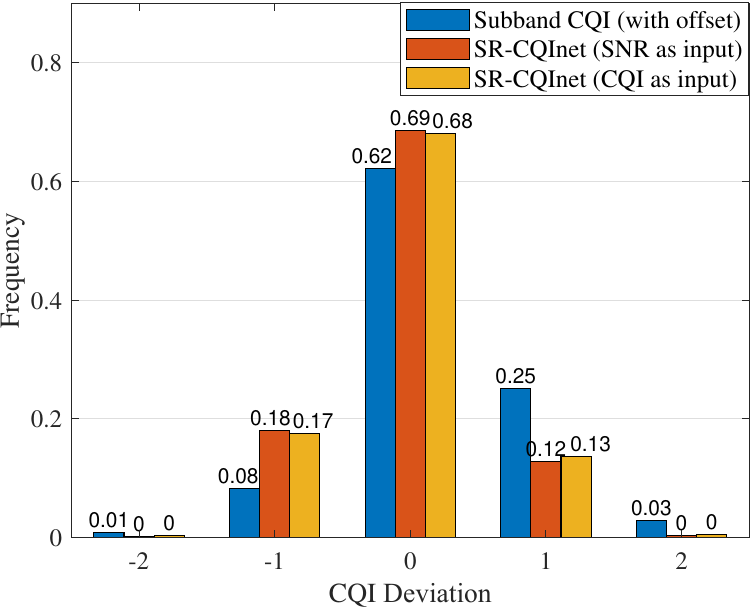}
        \caption{Comparison of CQI deviation distribution between SR-CQInet and the subband CQI scheme with offset.}
        \label{diff}
\end{figure}

\begin{figure}[t]
  \centering
  \includegraphics[width=0.9\linewidth]{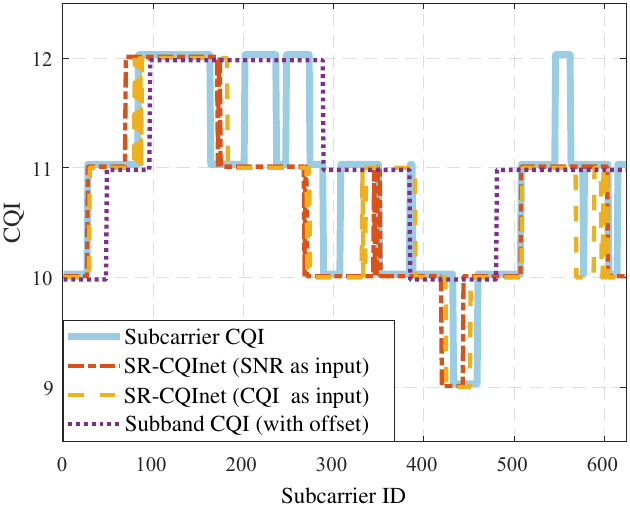}
        \caption{Sample comparison of SR-CQInet, the subband CQI scheme with offset, and the ground-truth subcarrier CQI.}
        \label{SRsubband}

\end{figure}


\begin{figure}[t]
  \centering
  \includegraphics[width=0.9\linewidth]{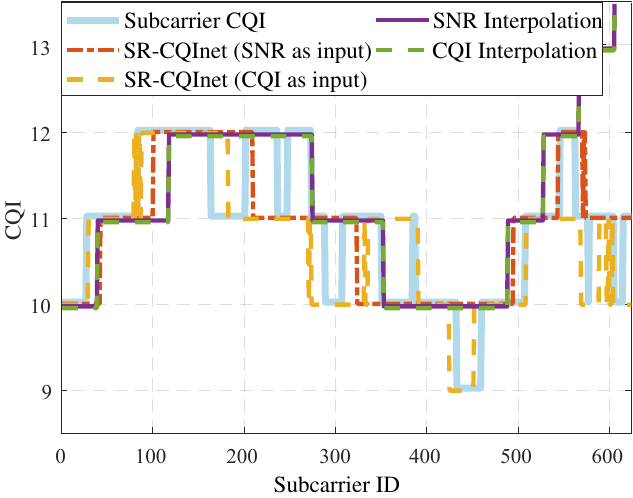}
        \caption{Sample comparison of SR-CQInet and linterpolation-based SR methods with ground-truth subcarrier CQI.}
        \label{SRinter}

\end{figure}

\begin{figure}[t]
  \centering
  \includegraphics[width=0.9\linewidth]{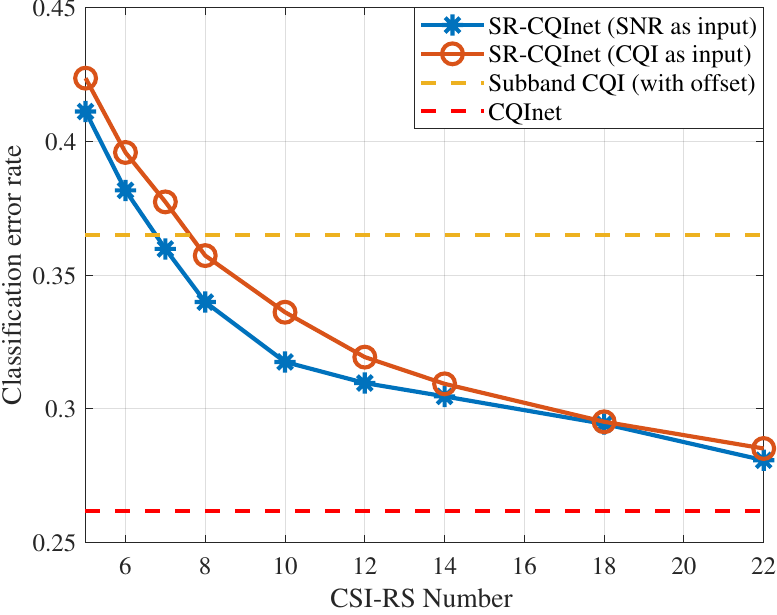}
        \caption{Comparison of classification error rates between SR-CQInet, CQInet, and the subband CQI scheme.}
        \label{classerror}

\end{figure}

\begin{figure}[t]
  \centering
  \includegraphics[width=0.9\linewidth]{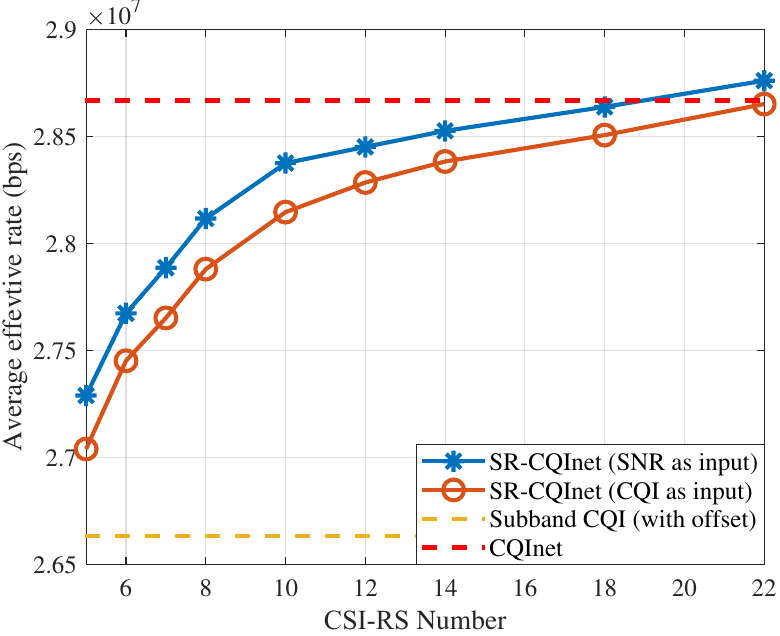}
        \caption{Comparison of effective rates between SR-CQInet, CQInet, and the subband CQI scheme.}
        \label{erate}

\end{figure}

We first evaluate the CQI selection accuracy of SR-CQInet by analyzing its deviation distribution under sparse CSI-RS deployment.
Fig.~\ref{diff} compares the CQI deviation distribution of SR-CQInet against the subband CQI scheme (with offset) when the number of CSI-RS deployed is ${N_{\rm CG}} = 13$. For a fair comparison, the feedback overhead $S$ is fixed at 30 bits for both schemes. The CQI deviation is defined as the difference between the reconstructed and ground-truth subcarrier CQI, where positive and negative values represent overestimation and underestimation, respectively.  
Results show that SR-CQInet, with either SNR or CQI input, significantly reduces the probability of one-level overestimation to 0.12 and 0.13, respectively, while increasing the probability of correct CQI estimation to 0.69 and 0.68. Additionally, it avoids severe misestimations beyond one level. These gains stem from SR-CQInet’s ability to exploit spatial correlations and leverage loss functions that penalize large deviations, thus improving link robustness and MCS reliability.
  
To further illustrate its performance, Fig.~\ref{SRsubband} presents a sample comparison between SR-CQInet and the subband CQI scheme (with offset) under the same configuration (${N_{\rm CG}} = 13$, $S = 30$).  
In the subcarrier range from the 430th to 460th, the subband CQI scheme reports a value of 10, whereas the true subcarrier CQI is 9. This overestimation risks high BLER due to overly aggressive MCS selection. In contrast, SR-CQInet (with both input types) accurately learns the local channel degradation and selects the correct CQI level of 9.  
Likewise, in the range from the 337th to 384th subcarriers, the subband CQI remains at 11 despite deep fades, while SR-CQInet captures these variations and, due to overestimation penalties in training, conservatively selects a slightly lower CQI to enhance reliability.


Moreover, to highlight the strength of NN-based SR, Fig.~\ref{SRinter} presents a sample comparison between SR-CQInet and a linear interpolation-based SR method. In this baseline, coarse-grained CQI is first reconstructed via an autoencoder with the same structure as CQInet and then upsampled using linear interpolation. 
Under ${N_{\rm CG}} = 8$ and $S = 30$, SR-CQInet successfully models the channel dip using surrounding context and assigns a CQI one level lower between the 430th and 460th subcarriers, closely matching ground truth. However, the interpolation method, with its narrow view, fails to capture this and assigns an inaccurate CQI.  
From the 570th to 624th subcarriers, where channel quality deteriorates due to frequency-selective fading, SR-CQInet models this slope and predicts a gradually decreasing CQI trend. In contrast, linear interpolation, limited by boundary information, predicts incorrectly, causing cumulative errors. These findings underscore SR-CQInet’s superiority in modeling non-linear channel features and performing high-resolution CQI estimation.
 
To evaluate system-level impact, Figs.~\ref{classerror} and \ref{erate} compare the CQI classification error and effective data rate of SR-CQInet, CQInet, and the subband CQI scheme (with offset), across different CSI-RS densities. Note that CQInet can be viewed as a special case of SR-CQInet with CQI input and ${N_{\rm CG}} = 624$ (i.e., full CSI-RS). Results show that when using fewer than 8 CSI-RS ports, SR-CQInet achieves better CQI accuracy and higher average throughput than the subband CQI scheme, reducing CSI-RS overhead by 38.5\%. These improvements validate SR-CQInet’s suitability for low-overhead, high-efficiency CQI feedback in 6G and NextG networks.

As CSI-RS density increases, SR-CQInet continues to improve in throughput, though with diminishing returns. For instance, with only 22 CSI-RS, it already matches and slightly exceeds the performance of CQInet when using SNR as input, even as its classification error rate remains approximately 2\% higher than CQInet's. This combination of results can be attributed to two core factors. First, CQInet's input (discrete CQI values for all 624 subcarriers) has higher dimensionality; while complete, it causes greater compression loss and higher training complexity, making it hard to fully preserve fine-grained channel features despite leveraging subcarrier intra-band spatial correlations. In contrast, SR-CQInet uses continuous SNR as input, which retains richer channel details, allowing its loss function to better prioritize suppressing overestimation. Second, CQI values are often stable across adjacent subcarriers. Thus, once a sufficient number of CSI-RS are available, critical variations are captured, and sparse deployment becomes effective.
As a result, SR-CQInet can operate with just 3.5\% of the CSI-RS overhead required by CQInet, making it an effective solution for practical systems where the CSI-RS count is limited.

Moreover, comparing input types reveals that SR-CQInet performs better with SNR input. On average, it reduces CQI error by 2\% and improves the effective rate by 1\% over CQI input. This advantage stems from SNR’s continuous representation of channel conditions, allowing SR-CQInet to better model fine-grained patterns.  
CQI, being quantized, lacks such granularity. Therefore, SNR input enables more accurate CQI prediction. Additionally, this suggests that the UE may directly report SNR values to the BS, bypassing CQI derivation and BLER-eSNR table lookups, thus simplifying the UE-side processing and reducing computational load.

\section{Conclusion}
\label{s5}

In this paper, we proposed a novel subcarrier-level CQI feedback framework that departs from the traditional subband-based approach by assigning CQI values to individual subcarriers. To address the resulting increase in feedback and reference signal overhead, we introduced deep learning-based solutions that enable efficient CQI compression and reconstruction, as well as SR estimation.
First, we developed a fine-grained subcarrier CQI scheme that improves channel adaptability and MCS selection granularity by bypassing the coarse eSNR mapping process used in subband CQI.  
To mitigate the increased feedback overhead, we designed \textbf{CQInet}, an autoencoder-based feedback algorithm that compresses and reconstructs subcarrier CQI under dense CSI-RS deployment.  
Furthermore, to address practical constraints on CSI-RS resources, we proposed \textbf{SR-CQInet}, which leverages SR techniques to reconstruct fine-grained CQI from sparse coarse-grained SNR or CQI inputs, significantly reducing CSI-RS overhead.

Simulation results demonstrated the effectiveness of our approach. The proposed subcarrier CQI scheme achieves a 13.2\% improvement in effective data rate compared to the conventional subband CQI scheme. CQInet further improves the effective rate by 7.6\% under the same feedback overhead as the traditional method.  
More importantly, SR-CQInet reduces CSI-RS overhead by 38.5\% while maintaining comparable performance, and it requires only 3.5\% of the CSI-RS overhead used by CQInet. This makes SR-CQInet particularly suitable for practical deployment of low-overhead, high-accuracy CQI feedback in 6G and NextG systems, where signaling efficiency and spectral utilization are paramount.


\bibliographystyle{IEEEtran}
\bibliography{IEEEabrv,reference}

\end{document}